\documentstyle[pra,aps,amssymb,amstex,twocolumn]{revtex}

\begin{document}

\draft

\title{Anomalous condensate fluctuations in strongly 
interacting superfluids}
\author{F. Meier and W. Zwerger}
\address{ Sektion Physik, Universit\"at M\"unchen, Theresienstrasse 
37, 
D-80333 M\"unchen, Germany }
\date{\today}

\maketitle

\begin{abstract}
We show that the condensate occupation of a superfluid Bose liquid 
quite generally exhibits anomalously large fluctuations at finite 
temperatures $T\not= 0$. In three dimensions, the variance $\langle \delta 
\hat{N}_0^2 \rangle$ of the number $\hat{N}_0$ of particles in the 
condensate scales nonlinearly with volume $V$ like $T^2 V^{4/3}$ at low $T$, 
generalizing the result obtained by Giorgini, Pitaevskii, and 
Stringari for a weakly interacting Bose gas. In two 
dimensions there is only a quasicondensate whose 
fluctuations are of the same order as the mean value.
\end{abstract}


The experimental realization of Bose Einstein condensation (BEC) in 
alkali vapours has stimulated extensive theoretical
research on the physics of BEC \cite{dgps99}. It is well 
known that the standard grand canonical ensemble used in textbooks 
gives unphysically large condensate fluctuations, $\langle \delta 
\hat{N}_0^2 \rangle \simeq \langle \hat{N}_0 \rangle^2 \propto V^2$, 
even at $T=0$. Much recent theoretical work has thus 
been devoted to the investigation of fluctuations in different 
statistical ensembles~\cite{politzer,navez,weiss,holthaus}. With a fixed total 
particle number, the pathologies of the grand canonical ensemble are 
removed, however the condensate fluctuations remain anomalously large, 
$\langle \delta \hat{N}_0^2 \rangle \propto T^2 V^{4/3}$, at least 
for an ideal Bose gas in a box. Exploiting the fact that BEC in a 
noninteracting gas is described exactly by the spherical model, 
this was first realized by Fierz \cite{fierz}. 
Surprisingly, the same behaviour was found to persist
in the presence of weak interactions as recently shown by
Giorgini, Pitaevski and Stringari using the Bogoliuobov theory
\cite{giorgini98}.

In this paper, we demonstrate that the anomalous condensate 
fluctuations are a general feature of any superfluid Bose system,  
valid for arbitrary interactions. The large fluctuations at finite 
temperature will be shown to be a direct consequence of Bogoliubov's 
well known $1/k^2$-theorem for the static susceptibility 
$\chi_{\varphi \varphi} (\vec{k})$ of superfluids~\cite{forster}.  
They are thus related to the existence of gapless modes and 
correspondingly are absent in artificial models with a finite gap in 
the spectrum as introduced recently in this context~\cite{wilkens}.
We give a very simple derivation similar to the one employed to prove 
that the thermal depletion of the condensate in dimension 
$d=3$ is $\propto T^2$~\cite{ferrell67}. For a Bose liquid in $d=2$ 
there is only a quasicondensate whose root mean square fluctuations
$\langle \delta \hat{N}_0^2 \rangle ^{1/2}$ are of the same order
than the mean value $\langle \hat{N}_{0} \rangle$, both scaling with 
volume with a temperature dependent exponent $1-\eta(T)/2$. 
Being a rather general property of a Bose condensed system, the
anomalous scaling with volume or particle number will also be
present for BEC's in harmonic traps, where the condensate
fluctuations are crucial for determining the linewidth of an 
atom laser \cite{bloch99,phillips99}
as recently shown by Graham \cite{graham98}. 

We restrict ourselves to Bose liquids in a fixed volume $V_d = L^d$ 
with periodic boundary conditions. As shown by Feynman, the spectrum 
of excitations as $k \rightarrow 0$ is exhausted by phononlike 
modes with $\omega_k = c k$. For low enough temperatures such that  
$mc^2 \gg k_B T$, only phase fluctuations are relevant~\cite{popov}
and the field operator may be written as~\cite{ll9}
\begin{equation}
\hat{\Psi}(\vec{x}) = \sqrt{n_0} \; e^{i \hat{\varphi}(\vec{x})}
\label{eq:fieldop}
\end{equation}
with $n_0$ being the bare condensate density at $T=0$ which would be 
observed in a small sample, i.e. in the absence of low energy 
excitations. The operator of phase fluctuations is
\begin{equation}
\hat{\varphi}(\vec{x}) = \sum_{\vec{k}} \mbox{}^{\prime} \left( 
\frac{m c}{2 V_d n \hbar k} \right)^{1/2} \left( \hat{c}_{\vec{k}} 
e^{i \vec{k} \cdot \vec{x}} + \hat{c}^\dagger_{\vec{k}} e^{-i \vec{k} 
\cdot \vec{x}}\right),
\label{eq:phaseop}
\end{equation}
where $m$ is the particle mass, $c$ the actual velocity of sound, and 
$n$ the mean particle density. The operators 
$\hat{c}^\dagger_{\vec{k}}$ and $\hat{c}_{\vec{k}}$ are phonon 
creation and annihilation operators, respectively. The prime 
indicates summation over all $\vec{k} \neq \vec{0}$ with $|\vec{k}| < 
\Lambda$, $\Lambda$ being a momentum cutoff (the term $k=0$ is
omitted because it corresponds to an irrelevant global phase factor
which can always be absorbed in the bare condensate density $n_{0}$
and drops out in any gauge invariant quantity as calculated below). 
The actual, renormalized 
condensate density is obtained from the off diagonal elements of the 
one particle density matrix, which gives
\begin{equation}
\lim_{|\vec{x}| \rightarrow \infty} \langle \hat{\Psi}^\dagger 
(\vec{x}) \Psi (\vec{0}) \rangle = n_0 \, \lim_{|\vec{x}| \rightarrow 
\infty} e^{-\langle 
(\hat{\varphi} (\vec{x}) - \hat{\varphi} (\vec{0}))^2 \rangle/2}
\label{eq:n0-1}
\end{equation}
as long as phonon interactions and higher energy excitations are 
neglected. For large systems, $\lim_{|\vec{x}| \rightarrow \infty} 
\langle \hat{\varphi} (\vec{x})  \hat{\varphi} (\vec{0}) \rangle$ 
is always small compared to $\langle \hat{\varphi}^2 (\vec{x}) 
\rangle$, which is independent of $\vec{x}$. Separating off the 
quantum fluctuations of $\hat{\varphi}$ at $T=0$, the mean condensate 
density at temperature $T$ may be written as
\begin{equation}
n_0(T) = n_0(T=0) \, e^{-(\left. \langle 
\hat{\varphi}^2(\vec{0})\rangle \right|_{T} -  \left. \langle 
\hat{\varphi}^2(\vec{0})\rangle \right|_{T=0}) }.
\label{eq:n0-2}
\end{equation}
In a strongly interacting Bose liquid, the condensate density at zero
temperature may be much smaller than the total density $n$, as is the 
case in superfluid \mbox{$^4$He}, where $n_0(T=0)/n \simeq 0.1$
\cite{ceperley95}.

In $d=3$, the exponent in (\ref{eq:n0-2}) is independent of the 
system size and proportional to $T^2$, thus describing the well known
thermal depletion of the condensate~\cite{ferrell67} 
\begin{equation}
\frac{n_0(T) - n_0(0)}{n_0(0)} = - \frac{m (k_B T)^2}{12 n c \hbar^3}.
\label{eq:depl}
\end{equation}

In $d=2$, the thermal contribution to phase fluctuations 
diverges logarithmically with system size~\cite{kane}
\begin{eqnarray}
& \left. \langle \hat{\varphi}^2(\vec{0})\rangle \right|_{T} -  
\left. \langle \hat{\varphi}^2(\vec{0})\rangle\right|_{T=0} & \simeq 
\frac{m k_B T}{2 \pi n \hbar^2} \ln (\Lambda L) \nonumber \\ && 
\equiv \eta(T) \, \ln (\Lambda L).
\label{eq:phasefluct}
\end{eqnarray}
Hence, the condensate density at finite temperature depends on the 
system size via $n_0(T) = n_0(0) \, (\Lambda L)^{-\eta (T)}$, with 
$\eta (T) \propto T$. In particular, for any finite $T$ the 
condensate density vanishes in the thermodynamic limit, reflecting the
fact that there is no true condensate in $d=2$ in agreement with the 
rigorous proof given by Hohenberg~\cite{hohenberg67}.

The operator of fluctuations around the mean field is given by
\begin{equation}
\delta \hat{\Psi}(\vec{x}) = \hat{\Psi} (\vec{x}) - \sqrt{n_0 (T)} = 
\sqrt{n_0} \, \left( e^{i \hat{\varphi} (\vec{x})} - e^{-\langle 
\hat{\varphi}^2(\vec{0}) \rangle /2}\right).
\label{eq:fluctop}
\end{equation}
In terms of $\delta \hat{\Psi} (\vec{x})$, the number of particles 
out of the condensate is
\begin{equation}
\hat{N}_{\text{out}} = \int d \vec{x} \; \delta \hat{\Psi}^\dagger 
(\vec{x}) \, \delta \hat{\Psi} (\vec{x}).
\label{eq:nout}
\end{equation}

Since the total number of particles is fixed, the fluctuations of 
the condensate occupation $\hat{N}_0$ may be calculated from $\langle 
\delta \hat{N}_0^2 \rangle = \langle \delta \hat{N}_{\text{out}}^2 
\rangle$, i.e. the condensate and noncondensate particles act as 
effectively infinite particle reservoirs for each 
other~\cite{politzer,fierz}. Using (\ref{eq:fluctop}) one obtains
\begin{equation}
\langle \delta \hat{N}_0^2 \rangle \simeq 2 n_0^2 e^{- 2 \langle 
\hat{\varphi}^2 (\vec{0}) \rangle} \int d \vec{x} \, d \vec{x^\prime} 
\, \langle \hat{\varphi} (\vec{x}) \hat{\varphi} (\vec{x^\prime}) 
\rangle^2,
\label{eq:fluct1}
\end{equation}
since $\langle \hat{\varphi} (\vec{x}) \hat{\varphi} (\vec{x^\prime}) 
\rangle$ is small at large distances both in $d=2$ and in $d=3$, 
provided that the temperature is low. Evaluating the correlation 
function at $T=0$, we find
\begin{eqnarray}
&& \left. \langle \delta \hat{N}_0^2 \rangle \right|_{T=0} \simeq
2 \left(\frac{n_0 (T=0)}{n} \right)^2 \left( \frac{m c}{2 \hbar} 
\right)^2  \nonumber \\ 
&& \hspace{2.5cm} \times \left\{ 
\begin{array}{ll}
\frac{1}{2 \pi^2} \Lambda V_3 & \text{d=3} \\ & \\
\frac{1}{4 \pi} V_2 \ln (\Lambda^2 V_2) & \text{d=2}.
\end{array}
\right.
\label{eq:fluct2}
\end{eqnarray}
The fluctuations are thus normal, scaling essentially linearly with 
the volume of the system.

At finite temperatures, it is convenient to express the integral in 
(\ref{eq:fluct1}) in terms of the static susceptibility $\chi_{\varphi 
\varphi}(\vec{k})$. Using the classical version of the fluctuation 
dissipation theorem, which is always applicable at low $k$ because 
$\hbar \omega_k \ll k_B T$, we find
\begin{equation}
\int d \vec{x} \, d \vec{x^{\prime}} \, \langle \hat{\varphi} 
(\vec{x}) \hat{\varphi} (\vec{x^{\prime}}) \rangle^2 = \sum_{\vec{k}} 
\mbox{}^{\prime} (k_B T \, \chi_{\varphi \varphi} (\vec{k}))^2.
\label{eq:fluct3}
\end{equation}
Since the superfluid density $n_{s}$ which usually 
appears in $\chi_{\varphi \varphi}$ \cite{forster} coincides with the
full density $n$ at zero temperature for any translation invariant 
superfluid, we have $\chi_{\varphi \varphi}(\vec{k}) = m/ n 
\hbar^2 k^2$ for $k \rightarrow 0$, independent of the interaction 
strength.  Hence, at low temperatures
\begin{equation}
\left. \langle \delta \hat{N}_0^2 \rangle \right|_T = 2 
\left(\frac{n_0(0)}{n} \right)^2 \left( \frac{m k_B T}{\hbar^2} 
\right)^2 \sum_{\vec{k}} \mbox{}^{\prime} \frac{1}{k^4}.
\label{eq:fluct4}
\end{equation}
For the weakly interacting Bose gas there is no depletion of
the condensate at $T=0$ to lowest order and thus with
$n_{0}(0)/n$ replaced by one, our result 
reduces to the one given in~\cite{giorgini98}. The somewhat surprising
fact that the velocity of sound does not appear in (12), may be
traced back to the well known low $k$-divergence \cite{ll9,hohenberg}
\begin{equation}
\lim_{k\to 0}n_{k}=\frac{n_{0}(0)}{n}\,\frac{k_{B}T}{2\varepsilon_{k}}
\label{eq:fluct5}
\end{equation}
of the momentum distribution at low but finite temperatures which,
apart from the ratio $n_{0}(0)/n$, only depends on the bare single particle
energy $\varepsilon_{k}=(\hbar k)^{2}/2m$. In order to make this 
connection more explicit, we note that the large distance behaviour of
the phase correlation function which appears in $(9)$ is directly 
related to the momentum distribution $n_{k}$ of the noncondensed particles
at low $k$ via~\cite{ll9}
\begin{eqnarray}
& \lim_{|\vec{x}| \rightarrow \infty} & \langle \hat{\Psi}^\dagger 
(\vec{x}) \Psi (\vec{0}) \rangle =  n_0(T)\left( 1+
\langle \hat{\varphi} (\vec{x}) \hat{\varphi} (0) \rangle\right)
\nonumber \\ &&
=n_{0}(T)\, +\int\,\frac{d^{3}k}{(2\pi)^{3}}\, n_{k}\exp{-i\vec k\vec 
x}
\end{eqnarray}
This shows that equ. $(9)$ can be written in the simple form
\begin{equation}
\langle \delta \hat{N}_0^2 \rangle =2\sum_{\vec{k}} 
\mbox{}^{\prime}\, n_{k}^{2}
\end{equation}
which immediately gives $(12)$ using the result $(13)$ for the
momentum distribution at finite temperature. The anomalous scaling of 
the condensate fluctuations with volume is therefore a direct 
consequence of the $1/k^{2}$-divergence of the momentum distribution
or - equivalently - the related slow $1/r$-decay of the one particle
density matrix at large distances to its limiting nonzero value $n_{0}(T)$.   
Incidentally this argument also 
explains why the result for an {\it ideal} Bose gas, where $n_{0}(0)=n$
trivially, is just twice the result (12). Indeed the 
momentum distribution for noninteracting Bosons at small nonzero $k$
is $n_{k}^{(0)}\to k_{B}T/\varepsilon_{k}$ (the factor two difference compared
with (13) is due to the fact that at low $k$ the quasiparticles of the
interacting system are just an equal weight superposition of
a bare particle with momentum $k$ plus a bare hole with momentum $-k$ 
\cite{nozieres}). Using the standard result for the occupation number
fluctuations of ideal Bosons we have
\begin{equation}
\langle \delta \hat{N}_0^2 \rangle^{(0)} =\sum_{\vec{k}} 
\mbox{}^{\prime}\, n_{k}^{(0)} ( 1+n_{k}^{(0)} ) \approx
\sum_{\vec{k}} \mbox{}^{\prime}\, ( n_{k}^{(0)} )^{2}.
\end{equation}
To leading order the fluctuations in the interacting and the 
noninteracting case are thus simply related by a factor
$2\left( n_{0}(0)/2n\right)^{2}$.
Contrary to the argument in \cite{giorgini98}, the
similar behaviour of the fluctuations in an ideal and interacting
Bose gas is therefore not accidental. 
 
As implied by 
(\ref{eq:fluct4}), at $T \neq 0$ the fluctuations of the ground state 
occupation are anomalous for any interaction. Expressing the result in
terms of the standard thermal wavelength $\lambda_{T}=h(2\pi 
mk_{B}T)^{-1/2}$ we have
\begin{equation}
\left. \langle \delta \hat{N}_0^2 \rangle \right|_T = B \left( 
\frac{n_0(0)}{n} \right)^2 
\left( \frac{L}{\lambda_{T}} \right)^4 \propto  V_3^{4/3}
\label{eq:fluct6}
\end{equation}
with a universal constant $B=\sum_{\vec{n}} \mbox{}^{\prime} 
|\vec n |^{-4}/2\pi^{2}=0.8375$ whose numerical value 
is readily obtained from the lattice sum (\ref{eq:fluct4})
(this result is roughly a factor five smaller than the corresponding 
one given in ref.\cite{giorgini98}).
Note that contributions $\vec k= 2\pi\vec n /L$ with one or two
of the components of $\vec n\in\mathbb{Z}$$^{3}$ being zero should be
included in the sum in the case of periodic boundary conditions as
assumed here. For Dirichlet boundary conditions in a box with hard 
walls the corresponding constant in an ideal Bose gas is in fact 
smaller by a factor $0.864$ \cite{martin}.

In $d=2$, the one particle density matrix decays to zero like
$1/r^{\eta(T)}$ at any finite temperature and the associated 
momentum distribution $n_{k}$ is thus proportional to $k^{-(2-\eta)}$
at small $k$. The resulting scaling of the condensate fluctuations 
with system size is temperature dependent 
\begin{eqnarray}
& \left. \langle \delta \hat{N}_0^2 \rangle \right|_T & \propto 
\left( \frac{n_0(0)}{n} \right)^2 
\left( \frac{L}{\lambda_{T}} \right)^4 \, (\Lambda L)^{-2\eta (T)}
\nonumber \\ && \propto V_2^{2-\eta(T)}.
\label{eq:fluct7}
\end{eqnarray}
increasing like the square of the average condensate number
$\langle \delta \hat{N}_0^2 \rangle \propto \langle \hat{N}_0 
\rangle^2$. The fluctuations in $d=2$ are 
thus much larger than those in $d=3$. The fact 
that the rms fluctuations of the condensate are of the same 
order as the average is yet 
another indication that the condensate in $d=2$ is not well defined. 

In conclusion, we have shown that the anomalous 
fluctuations of the condensate occupation at finite temperature, 
found previously for both ideal and weakly interacting Bose gases, 
are a general property of any superfluid with arbitrary
interactions. . The anomalous behaviour of these fluctuations 
may be traced back via Bogoliubov's $1/k^2$-theorem to the
corresponding low $k$-singularity in the momentum distribution.
As has been shown in \cite{giorgini98}, the condensate fluctuations 
in harmonic traps are similarly determined by the 
associated low lying collective 
excitations. As a result, the anomalous scaling 
$\left. \langle \delta \hat{N}_0^2 \rangle \right|_T \propto 
T^{2} N^{4/3}$ with the particle number $N$ 
is also expected to be valid quite generally

\acknowledgements{It is a pleasure to thank M. Holthaus, G. Shlyapnikov 
and C. Weiss for helpful discussions and remarks on the manuscript.}

\end{document}